\documentclass[dvips,12pt,a4paper]{article}

\usepackage[greek,american]{babel}
\usepackage{amssymb}
\usepackage{amsfonts}
\usepackage{amsmath}
\usepackage{relsize}
\usepackage[lofdepth,lotdepth]{subfig}

\usepackage{bm}
\usepackage[dvips]{graphicx}

\usepackage{float}
\usepackage[
singlelinecheck=false 
]{caption}

\renewcommand{\vec}[1]{\mathbf{#1}}

\usepackage{amsthm}

\usepackage{mathtools}

\usepackage{algorithm}
\usepackage{algpseudocode}
\usepackage{pifont}

\usepackage{bbm}

\def\P{\mathbb P}
\def\G{\mathbb G}
\def\R{\mathbb R}
\def\E{\mathbb E}
\def\Q{\mathbb Q}
\def\N{\mathbb N}
\def\d{\delta}
\def\t{\theta}
\def\l{\lambda}
\def\a{\alpha}
\def\b{\beta}

\def\iid{\stackrel{\rm iid}\sim}
\def\ind{\stackrel{\rm ind}\sim}

\graphicspath{{art/}}

\begin{document}

\begin{center}

{{{\Large\sf\bf Dependent Mixtures of Geometric Weights Priors}}}\\

\vspace{0.5cm}
{\sf Spyridon J. Hatjispyros
     \footnote{Corresponding author. Tel.:+30 22730 82326\\
     \indent E-mail address: schatz@aegean.gr }$^{,\,*}$,
      Christos Merkatas$^{*}$,
      Theodoros Nicoleris$^{**}$,
      Stephen G. Walker$^{***}$}

\vspace{0.2cm}
\end{center}
\centerline{\sf $^{*}$ Department of Mathematics, University of the Aegean,}
\centerline {\sf Karlovassi, Samos, GR-832 00, Greece.} 
\centerline{\sf $^{**}$ Department of Economics, National and Kapodistrian University of Athens,}
\centerline{\sf Athens, GR-105 59, Greece. }  
\centerline{\sf $^{***}$Department of Mathematics, University of Texas at Austin,}
\centerline{\sf Austin, Texas 7812, USA. }

\begin{abstract}
A new approach to the joint estimation of partially exchangeable observations is presented. This is achieved by constructing 
a model with pairwise dependence between random density functions, each of which is modeled as a mixture of \textit{geometric} 
stick breaking processes. 
The claim is that mixture modeling with Pairwise Dependent Geometric Stick 
Breaking Process (PDGSBP) priors is sufficient for prediction and estimation purposes; that is, making the weights more exotic does not actually enlarge the support of the prior. Moreover, the corresponding Gibbs sampler for estimation
is faster and easier to implement than the Dirichlet Process counterpart. 

\vspace{0.1in} \noindent 
\end{abstract}

\vspace{0.5in} \noindent {\bf 1. Introduction.} 
In Bayesian nonparametric methods, the use of priors such as the Dirichlet process (Ferguson, 1973),  is justified from the assumption 
that the observations are exchangeable, which means the distribution of $(X_{1},\ldots,X_{n})$ 
coincides with the distribution of $(X_{\pi(1)},\ldots,X_{\pi(n)}),$ 
for all $\pi\in S(n)$, where $S(n)$ is the set of permutations of  $\{1,\ldots,n\}$. 
However, in real life applications, data are often \textit{partially exchangeable}. 
For example, they may consist of observations sampled from $m$ populations, 
or may be sampled from an experiment conducted in $m$ different geographical areas.
This means that the joint law is invariant under permutations within the $m$ subgroups of observations
$(X_{j,i_j})_{1\le i_j\le n_j},\,1\le j\le m$, so for all $\pi_j\in S(n_j)$
{
\begin{equation}
\label{Partially1}
((X_{1,i_1})_{1\le i_1\le n_1},\ldots,(X_{m, i_m})_{1\le i_m\le n_m})\sim
((X_{1,\pi_1(i_1)})_{1\le i_1\le n_1},\ldots,(X_{m, \pi_m(i_m)})_{1\le i_m\le n_m}).
\end{equation}
}
When the exchangeability assumption fails one needs to use non--exchangeable priors. There has been substantial research interest 
following the seminal work of MacEachern (1999) in the construction of suitable dependent stochastic processes.  Such then act as priors in 
Bayesian nonparametric models. These processes are distributions over a collection of measures indexed by values in some covariate 
space, such that the marginal distribution is described by a known nonparametric prior. The key idea is to induce dependence between 
a collection of random  probability measures $(\P_j)_{1\le j\le m}$, where each $\P_j$ comes from a Dirichlet process (DP) with concentration 
parameter $c>0$ and base measure $P_0$. Such random probability measures typically are used in mixture models to generate random density functions $f(x) = \int_{\Theta}K(x|\theta)\P(d\theta)$; see Lo (1984).

There is a variety of ways that a DP can be extended to dependent DP. 
Most of them use the stick-breaking representation (Sethuraman, 1994), that is
$$
\P(\,\cdot\,) = \sum_{k=1}^{\infty}w_{k}\delta_{\theta_{k}}(\,\cdot\,),
$$
where $(\theta_k)_{k\ge 1}$ are independent and identically distributed from $P_{0}$ and $(w_k)_{k\ge 1}$ is
a stick breaking process; so if $(v_k)_{k\ge 1}$ are independent and identically distributed from 
${\cal B}e(1, c)$, a beta distribution with mean $(1+c)^{-1}$, then $w_{1}=v_{1}$ and for $k>1$, $w_{k}=v_{k}\prod_{l<k}(1-v_l)$.
Dependence is introduced through the weights and/or the atoms. 
A classical example of the use  of dependent DP's is the Bayesian nonparametric regression problem where a random probability measure 
$\P_{z}$ is constructed for each covariate $z$, 
$$
\P_z(\,\cdot\,) = \sum_{k=1}^\infty w_k(z)\delta_{\theta_k(z)}(\,\cdot\,),
$$
where $(w_{k}(z),\theta_{k}(z))$ is a collection of processes indexed in $z$--space.
Extensions to dependent DP models can be found in De Iorio et al. (2004), Griffin and Steel (2006), and Dunson and Park (2008). 

Recently there has been growing interest for the use of simpler random probability measures
which while simpler are  yet sufficient for Bayesian nonparametric density estimation. The geometric stick breaking 
(GSB) random probability measure (Fuentes--Garc\'ia, et al. 2010) has been used for density estimation and has been 
shown to provide an efficient alternative to DP mixture models. Some recent papers extend this 
nonparametric prior to a dependent nonparametric prior. In the direction of covariate dependent processes, GSB processes 
have been seen to provide an adequate model to the traditional dependent DP model. For example, for Bayesian 
regression, Fuentes--Garcia et al. (2009) propose a covariate dependent process based on random probability measures 
drawn from a GSB process. Mena et al. (2011) used GSB random probability measures in order to construct a purely atomic 
continuous time measure--valued process, useful for the analysis of time series data. In this case, the covariate $z\geq 0$  
denotes the time that each observation is (discretely) recorded and conditionally on each observation is drawn from a 
time--dependent nonparametric mixture model based on GSB processes. However, to the best of our knowledge, random probability 
measures drawn from a GSB process, for modeling related density functions when samples from each density function are available,
has not been developed in the literature.

In this paper we will construct pairwise dependent random probability measures based on GSB processes. 
That is, we are going to model a finite collection of $m$ random distribution functions $(\G_j)_{1\le j\le m}$, 
where each $\G_{j}$ is a GSB random probability measure, such that there is a unique common component for 
each pair $(\G_{j},\G_{j'})$ with $j\neq j'$. We are going to use these measures in the context of GSB mixture models, 
generating a collection of $m$ GSB pairwise dependent random densities $(f_{j}(x))_{1\le j\le m}$. 
Hence we obtain a set of random densities $(f_{1},\ldots,f_{m})$, 
where marginally each $f_{j}$ is a random density function
$$
f_{j}(x) = \int_{\Theta}K(x|\theta)\,\G_{j}(d\theta),
$$
thus generalizing the GSB priors to a multivariate setting for partially exchangeable observations.

In the problem considered here, these random density functions $(f_{j})_{1\le j\le m}$ are thought to be related or similar, e.g. perturbations of each other, and so we aim to share information between groups to improve estimation of each density, especially for those densities 
$f_{j}$ for which the corresponding sample size $n_{j}$ is small.  
In this direction, the main references include the work of M\"uller et al. (2014), Bulla et al. (2009),  
Kolossiatis et al. (2013) and Griffin et al. (2013); more rigorous results can be found in Lijoi et al.
(2014A, 2014B). 
All these models have been proposed for the modeling of an arbitrary but 
finite number of random distribution functions, via a common part and an index specific idiosyncratic part
so that for $0<p_j<1$ we have
$
\P_{j} = p_{j}\P_0 + (1-p_{j})\P_{j}^*,
$
where $\P_0$ is the common component to all other distributions and $\{\P_{j}^*:j=1,\ldots,m\}$ are the  
idiosyncratic parts to each $\P_{j}$, and $\P_0,\P_{j}^*\iid {\cal DP}(c,P_0)$. In Lijoi et al. (2014B)  
normalized random probability measures based on the $\sigma$--stable process are used for modeling 
dependent mixtures. Although similar (all models coincide only for the $m = 2$ case), these models are different 
from our model which is based on pairwise dependence of a sequence of random measures (Hatjispyros et al. 2011, 2016A). 

We are going to provide evidence through numerical experiments that dependent GSB mixture models are 
an efficient alternative to pairwise dependent DP (PDDP) priors. First, we will randomize the existing 
PDDP model of Hatjispyros et al. (2011, 2016A), by imposing gamma priors on the concentration masses (leading to
the more efficient rPDDP model). Then, for the objective comparison of execution times, we will  
conduct a-priori synchronized density estimation comparison studies between 
the randomized PDDP and the pairwise dependent GSB process (PDGSBP) models using synthetic and real data examples. 

This paper is organized as follows. In Section 2 we will demonstrate the construction of pairwise 
dependent random densities, using a dependent 
model suggested by Hatjispyros et al. (2011). We also demonstrate how specific choices of latent 
random variables can recover the model of Hatjispyros et al. and the dependent 
GSB model introduced in this paper. These latent variables will form the basis of a Gibbs sampler 
for posterior inference, given in Section 3. In Section 4 we resort to simulation. We provide 
comparison studies between the randomized version of the PDDP model and our newly introduced
dependent GSB model, involving five cases of synthetic data and a real data set.
Finally, Section 5 concludes with a summary and future work.


\vspace{0.2in} \noindent{\bf 2. Preliminaries.}
We consider an infinite real valued process $\{X_{ji}:1\le j\le m,\,i\ge 1\}$
defined over a probability space $(\Omega,{\cal F}, {\rm P})$, 
that is partially exchangeable as in (\ref{Partially1}). Let ${\cal P}$ denote
the set of probability measures over $\R$; then de Finetti proved that there exists
a probability distribution $\Pi$ over ${\cal P}^m$, which satisfies  
{
\begin{align}
 \nonumber 
   & {\rm P}\{X_{ji}\in A_{ji}:1\le j\le m,1\le i\le n_j\}\\
 \nonumber 
   & = \int_{{\cal P}^m}{\rm P}\{X_{ji}\in A_{ji}:1\le j\le m,1\le i\le n_j\,|\,\Q_1,\ldots,\Q_m\}\,\Pi(d\Q_1,\ldots,d\Q_m)\\
  \nonumber 
   & = \int_{{\cal P}^m}\prod_{j=1}^m{\rm P}\{X_{ji}\in A_{ji}:1\le i\le n_j\,|\,\Q_j\}\,\Pi(d\Q_1,\ldots,d\Q_m)\\
  \nonumber      
   &  = \int_{{\cal P}^m}\prod_{j=1}^m\,\prod_{i=1}^{n_j}\Q_j(A_{ji})\,\Pi(d\Q_1,\ldots,d\Q_m)\,.  
\end{align}
}%
The de Finetti measure $\Pi$ represents a prior distribution over partially exchangeable observations. 

We start off by describing the PDDP model, with no auxiliary variables, using only the 
de Finetti measure $\Pi$, marginal measures $\Q_j$, then, we proceed to the definition 
of a randomized version of it, and to the specific details for the case of the GSB random measures. 

\vspace{0.1in} \noindent {\bf A.}
In Hatjispyros et al. (2011), the following hierarchical 
model was introduced. For $m$ subgroups of observations 
$\{(x_{ji})_{1\le i\le n_j}:1\le j\le m\}$,  
\begin{align}
x_{ji}|\theta_{ji} & \ind K(\,\cdot\,|\theta_{ji})\nonumber\\
\theta_{ji}|\Q_j & \iid\Q_j(\,\cdot\,)\nonumber\\
\Q_j  = &\sum_{l=1}^mp_{jl}\P_{jl},\,\,\sum_{l=1}^mp_{jl}=1,\,\,\P_{jl}=\P_{lj}\nonumber\\
\P_{jl}   \iid & \,\,{\cal DP}(c, P_0),\,\,1\le j\le l\le m,\nonumber
\end{align}
for some kernel density $K(\,\cdot\,|\,\cdot\,)$, concentration parameter $c>0$ and parametric 
central measure $P_0$ for which $\E(\P_{jl}(d\theta))=P_0(d\theta)$.
So, we have assumed that the random densities $f_j(x)$ are dependent mixtures 
of the dependent random measures $\Q_j$ via $f_j(x|\Q_j)=\int_\Theta K(x|\theta)\Q_j(d\theta)$,
or equivalently, dependent mixtures of the $m$ independent mixtures 
$g_{jl}(x|\,\P_{jl})=\int_\Theta K(x|\,\t)\,\P_{jl}(d\t)$, $l=1,\ldots,m$.
To introduce the rPDDP model, we randomize the PDDP model by sampling the 
$\P_{jl}$ measures from the independent Dirichlet 
processes ${\cal DP}(c_{jl},P_0)$ and then impose gamma priors on the concentration masses, i.e.
$
\P_{jl} \ind \,{\cal DP}(c_{jl}, P_0),\quad c_{jl}\ind{\cal G}(a_{jl},b_{jl}),\,\,1\le j\le l\le m.
$

\vspace{0.1in} \noindent {\bf B.}
To develop a pairwise dependent geometric stick breaking version, 
we let the random density functions $f_j(x)$ generated via
\begin{equation}
\label{fjofx}
f_j(x):=f_j(x|\,\Q_j)\, = \,
\sum_{l=1}^m p_{jl}\,g_{j\,l}(x|\,\G_{jl}),\quad\Q_j  = \sum_{l=1}^mp_{jl}\G_{jl},\quad 1\le j\le m.
\end{equation}
The $g_{jl}(x):=g_{jl}(x|\,\G_{jl})=\int_\Theta K(x|\,\t)\,\G_{jl}(d\t)$ random densities are 
now independent mixtures of GSB processes, satisfying $g_{jl}=g_{lj}$, under the slightly altered definition
\begin{equation}
\label{abuseddef1}
\G_{jl}=\sum_{k=1}^\infty q_{jlk}\delta_{\theta_{jlk}}\quad{\rm with}\quad q_{jlk}=\lambda_{jl}(1-\lambda_{jl})^{k-1},\,\,
\lambda_{jl}\sim h(\,\cdot\,|\xi_{jl}),\,\,\,\theta_{jlk}\iid G_0,
\end{equation}
where $h$ is a parametric density supported over the interval $(0,1)$ depending on some parameter $\xi_{jl}\in\Xi$, 
and $G_0$ is the associated parametric central measure. 

The independent GSB processes $\{\G_{jl}:\,1\le j,\,l\le m\}$ form a matrix 
$\G$ of random distributions with $\G_{jl}=\G_{lj}$.
In matrix notation
\begin{equation}
\label{Qmeasurematrix}
{\mathbb Q}\,=\,\left(p\otimes \G\right){\bf 1},
\end{equation}
where $p=(p_{jl})$ is the matrix of random selection weights, and $p\otimes\G$ is the Hadamard
product of the two matrices defined as $(p\otimes\G)_{jl}=p_{jl}\G_{jl}$. By letting
$\bf 1$ to denote the $m\times 1$ matrix of ones it is that the $j$th element of vector 
$\mathbb Q$ is given by equation (\ref{fjofx}).

\vspace{0.1in} \noindent {\bf C.}
Following a univariate construction of geometric slice sets (Fuentes--Garc\'ia et al. 2010), we define
the stochastic variables ${\mathbf N}=(N_{ji})$ for $1\le i\le n_j$ and $1\le j\le m$, where $N_{ji}$ is an almost 
surely finite random variable of mass $f_{N}$ possibly depending 
on parameters, associated with the sequential slice set ${\cal S}_{ji}=\{1,\ldots,N_{ji}\}$. 
Following Hatjispyros et al. (2011, 2016a) we introduce: 
\begin{enumerate}

  \item The GSB mixture selection variables ${\boldsymbol\delta}=(\d_{ji})$; 
        for an observation $x_{ji}$ that comes from $f_j(x)$, 
        $\d_{ji}$ selects one of the mixtures $\{g_{jl}(x):l=1,\ldots,m\}$. 
        Then the observation $x_{ji}$ came from mixture $g_{j \delta_{ji}}(x)$.
        
  \item The GSB clustering variables ${\boldsymbol d}=(d_{ji})$; for an observation 
        $x_{ji}$ that comes from $f_j(x)$, given $\delta_{ji}$, $d_{ji}$ allocates the 
        component of the GSB mixture $g_{j \delta_{ji}}(x)$ that $x_{ji}$ came from. 
        Then the observation $x_{ji}$ came from component $K(x|\t_{j\d_{ji}d_{ji}})$.
        
\end{enumerate}

In what follows, unless otherwise specified, the random densities $f_j(x)$ are mixtures of independent GSB mixtures.

\vspace{0.2in} \noindent
{\bf Proposition 1.~}{\sl
Suppose that the clustering variables $(d_{ji})$ conditionally on the slice variables $(N_{ji})$ are having 
the discrete uniform distribution over the sets $({\cal S}_{ji})$ that is $d_{ji}|N_{ji}\sim{\cal DU}({\cal S}_{ji})$, and ${\rm P}\{N_{ji}=r|\d_{ji}=l\}=f_N(r|\lambda_{jl})$, then
\begin{equation}
\label{proposition11}
f_j(x_{ji},N_{ji}=r)=r^{-1}\sum_{l=1}^{m}p_{jl}f_{N}(r|\lambda_{jl})\sum_{k=1}^{r}\,K(x_{ji}|\theta_{jlk}),
\end{equation}
and 
\begin{equation}
\label{proposition12}
f_{j}(x_{ji},N_{ji}=r,d_{ji}=k|\delta_{ji}=l) = {1\over r}f_N(r|\lambda_{jl})\,{\cal I}(k\le r)\,K(x_{ji}|\theta_{jlk}).
\end{equation}
}%
\smallskip
{\sl The proof is given in Appendix A.}

\vspace{0.2in}
The following proposition gives a multivariate analogue of equation ($2$) in Fuentes--Garc\'ia, et al. (2010):

\vspace{0.2in} \noindent
{\bf Proposition 2.~}{\sl
Given the random set ${\cal S}_{ji}$, the random functions in (\ref{fjofx}) become finite mixtures of a.s. finite equally weighted mixtures 
of the $K(\,\cdot\,|\,\cdot\,)$ probability kernels, that is
\begin{equation}
\label{proposition21}
f_j(x_{ji}|N_{ji}=r)=\sum_{l=1}^m{\cal W}(r|\lambda_{jl})\sum_{k=1}^{r}r^{-1}K(x_{ji}|\theta_{jlk}),
\end{equation}
where the probability weights $\{ {\cal W}(r|\lambda_{jl}):1\le l\le m\}$ are given by
$$
{\cal W}(r|\lambda_{jl})={p_{jl} f_N(r|\lambda_{jl})\over \sum_{l'=1}^m p_{jl'} f_N(r|\lambda_{jl'})}.
$$
}%
\smallskip
{\sl The proof is given in Appendix A.}

Note that, the one--dimensional model introduced in Fuentes--Garc\'ia et al. (2010), under our notation attains the representation
$$
f_j(x_{ji}|N_{ji}=r,\d_{ji}=l)=\sum_{k=1}^{r}r^{-1}K(x_{ji}|\theta_{jlk}).
$$


\vspace{0.2in} \noindent{\bf 2.1 The model.} Marginalizing (\ref{proposition12}) with respect to the variable $(N_{ji}, d_{ji})$, we obtain
\begin{equation}
\label{fjgivendelta}
   f_j(x_{ji}|\d_{ji}=l)=\sum_{k=1}^\infty\left(\sum_{r=k}^\infty
    r^{-1}f_N(r|\lambda_{jl})\right)K(x_{ji}|\theta_{jlk}).
\end{equation}
The quantity inside the parentheses on the right-hand side of the previous equation is $f_j(d_{ji}|\d_{ji}=l)$.
Following Fuentes--Garc\'ia, et al. (2010), we substitute $f_N(r|\l_{jl})$ with the negative binomial distribution 
${\cal NB}(r|2,\l_{jl})$, i.e. 
\begin{equation}
\label{NB2}
f_N(r|\l_{jl})  = r \l_{jl}^2(1-\l_{jl})^{r-1}{\cal I}(r\ge 1),
\end{equation}
so equation (\ref{fjgivendelta}) becomes
$$
f_j(x_{ji}|\d_{ji}=l)=\sum_{k=1}^\infty q_{jlk} K(x_{ji}|\theta_{jlk})\,\,\,{\rm with}\,\,\,q_{jlk}=\l_{jl}(1-\l_{jl})^{k-1},
$$ 
and the $f_j$ random densities take the form of a finite mixture of GSB mixtures
$$
f_j(x_{ji})=\sum_{l=1}^m p_{jl}\sum_{k=1}^\infty q_{jlk} K(x_{ji}|\theta_{jlk}).
$$
We denote the set of observations along the $m$ groups as ${\boldsymbol x}=(x_{ji})$ and with 
${\boldsymbol x}_j$ the set of observations in the $j$th group.
The three sets of latent variables in the $j$th group will be denoted as ${\boldsymbol N}_j$ for the slice variables, 
${\boldsymbol d}_j$ for the clustering variables, and finally ${\boldsymbol \d}_j$ for the set of GSB mixture allocation variables. 
From now on, we are going to leave the auxiliary variables unspecified; especially for $\d_{ji}$ we use the notation
$
\d_{ji}=(\d_{ji}^1,\ldots,\d_{ji}^m)\in\left\{\vec{e}_1,\ldots,\vec{e}_m\right\}\,\,\,{\rm with}\,\,\,{\rm P}\{\d_{ji}=\vec{e}_l\}=p_{jl},
$
where $\vec{e}_l$ denotes the usual basis vector having its only nonzero component equal to $1$ at position $l$. 
Hence, for a sample of size $n_1$ from $f_1$, a sample of size $n_2$ from $f_2$, etc.,
a sample of size $n_m$ from $f_m$ we can write the full likelihood as a multiple product:
\begin{eqnarray}
f({\boldsymbol x},{\boldsymbol N},{\boldsymbol d}\,|\,{\boldsymbol \d}) 
       & = & \prod_{j=1}^m  f({\boldsymbol x}_j,{\boldsymbol N}_j,{\boldsymbol d}_j\,|\,{\boldsymbol \d}_j)\nonumber \\
       & = & \prod_{j=1}^m\prod_{i=1}^{n_j}{\cal I}(d_{ji}\le N_{ji})\prod_{l=1}^m
             \left\{\l_{jl}^2(1-\l_{jl})^{N_{ji}-1}K(x_{ji}|\,\theta_{j l d_{ji}})\right\}^{\delta_{ji}^l}.\nonumber
\end{eqnarray}
In a hierarchical fashion, using the auxiliary variables, we have for $j=1,\ldots,m \text{ and } i = 1,\ldots, n_{j},$
\begin{align}
\nonumber
    & x_{ji}, N_{ji}\,|\, d_{ji}, \d_{ji}, (\t_{jr\d_{ji}})_{1\le r\le m},\lambda_{j\d_{ji}} \,\ind\, 
      \prod_{r=1}^{m}\left\{\l_{jr}^2(1-\l_{jr})^{N_{ji}-1}K(x_{ji}|\theta_{jr d_{ji}})\right\}^{\delta_{ji}^{r}}{\cal I}(N_{ji}\ge d_{ji})\\
\nonumber
    & d_{ji}\,|\,N_{ji} \ind {\cal DU}({\cal S}_{ji}),\,\,\,{\rm P}\{\delta_{ji}=\vec{e}_{l}\} = p_{jl}\\
\nonumber
    & q_{jik}=\lambda_{ji}(1-\lambda_{ji})^{k-1},\,\,\,\t_{jik}\iid G_0,\,\,\,k\in\N.  
\end{align}


\vspace{0.2in} \noindent{\bf 2.2 The PDGSBP covariance and correlation.} In this sub--section we find the covariance and the correlation between $f_j(x)$ and $f_i(x)$.
First we provide the following lemma.

\vspace{0.2in} \noindent
{\bf Lemma 1.~}{\sl
Let $g_\G(x)=\int_\Theta K(x|\t)\G(d\t)$ be a random density, with 
$\G=\l\sum_{j=1}^\infty (1-\l)^{j-1}\d_{\t_j}$ and $\t_j\iid G_0$, then
$$
 \E[g_\G(x)^2] = \left({1\over 2-\l}\right)\left\{\l\int_\Theta K(x|\theta)^2G_{0}(d\theta) 
 + 2(1-\l)\left(\int_\Theta K(x|\theta)G_{0}(d\theta)\right)^2\right\}.
$$
}%
\smallskip
{\sl The proof is given in Appendix A.}

\vspace{0.2in} \noindent
{\bf Proposition 3.~}{\sl It is that
\begin{equation}
\label{CovPDSBP}
{\rm Cov}(f_j(x),f_i(x))\, =\, p_{ji}\,p_{ij}{\rm Var}\left(\int_\Theta K(x|\t)\G_{ji}(d\t)\right),
\end{equation}
with
\begin{equation}
\label{VarPDSBP}
{\rm Var}\left(\int_\Theta K(x|\t)\G_{ji}(d\t)\right)={\l_{ji}\over 2-\l_{ji}}{\rm Var}(K(x|\t)).
\end{equation}
}%
\smallskip
{\sl The proof is given in Appendix A.}

Suppose now that $(f_j^{\cal D}(x))_{1\le j\le m}$ and 
$(f_j^{\cal G}(x))_{1\le j\le m}$ are two collections
of $m$ DP and $m$ GSB pairwise dependent random densities respectively, i.e.
$f_j^{\cal D}(x)=\sum_{l=1}^m p_{jl}g_{jl}^{\cal D}(x)$ with $g_{jl}^{\cal D}(x)=g_{jl}(x|\P_{jl})$, and
$f_j^{\cal G}(x)=\sum_{l=1}^m p_{jl}g_{jl}^{\cal G}(x)$ with $g_{jl}^{\cal G}(x)=g_{jl}(x|\G_{jl})$.
Then we have the following proposition:

\vspace{0.2in} \noindent
{\bf Proposition 4.~}{\sl For given parameters $(\l_{ji})$, $(c_{ji})$, and matrix of selection
probabilities $(p_{ji})$ it is that
\begin{enumerate}
\item
The PDGSBP and rPDDP correlations are given by  
\begin{equation}
\label{PDGSBP_Corr}
{\rm Corr}(f_j^{\cal G}(x),f_i^{\cal G}(x))  = {\lambda_{ji}p_{ji}p_{ij}\over 2-\lambda_{ji}}
\left(\sum_{l=1}^m\sum_{r=1}^m  {p_{jl}^2 p_{ir}^2\lambda_{jl}\lambda_{ir}\over (2-\lambda_{jl})(2-\lambda_{ir})}\right)^{-1/2},
\end{equation}
and 
\begin{equation}
\label{rPDDP_Corr}
{\rm Corr}(f_j^{\cal D}(x),f_i^{\cal D}(x)) = {p_{ji}p_{ij}\over 1+c_{ji}}
\left(\sum_{l=1}^m\sum_{r=1}^m  {p_{jl}^2 p_{ir}^2\over (1+c_{jl})(1+c_{ir})}\right)^{-1/2}.
\end{equation}

\item 
When $\lambda_{ji}=\lambda$ and $c_{ji}=c$ for all $1\le j\le i\le m$, 
the expressions for the rPDDP and PDGSBP correlations simplify to
$$
{\rm Corr}(f_j^{\cal G}(x),f_i^{\cal G}(x))={\rm Corr}(f_j^{\cal D}(x),f_i^{\cal D}(x))
 = p_{ji}p_{ij}\left( \sum_{l=1}^m \sum_{r=1}^m  p_{jl}^2 p_{ir}^2 \right)^{-1/2}.
$$
\end{enumerate}
}%
\smallskip
{\sl The proof is given in Appendix A.}

\vspace{0.2in} 
It is clear that, irrespective of the model, the random densities $f_j(x)$ and $f_i(x)$
are positively correlated whenever $p_{ji}=p_{ij}=1$. Similarly, the random 
densities $f_j(x)$ and $f_i(x)$ are independent (have no common part) whenever $p_{ji}=p_{ij}=0$.  
Another, less obvious feature, upon synchronization, is the ability of controlling the 
correlation among the models. For example, suppose that for $m=2$, the random densities $f_1(x)$ 
and $f_2(x)$ are dependent, and that $\l_{ji}=(1+c_{ji})^{-1}$; then consider the expression
$$
D_{12}:=\l_{12}^2\, p_{12}^2\, p_{21}^2\,\left\{
{\rm Corr}(f_1^{\cal G}(x),f_2^{\cal G}(x))^{-2}
-{\rm Corr}(f_1^{\cal D}(x),f_2^{\cal D}(x))^{-2}\right\}.
$$
Since correlations are positive,
$D_{12}\ge 0$ whenever
${\rm Corr}(f_1^{\cal G}(x),f_2^{\cal G}(x))\le 
{\rm Corr}(f_1^{\cal D}(x),f_2^{\cal D}(x)),$ and that
$D_{12}<0$ whenever ${\rm Corr}(f_1^{\cal G}(x),f_2^{\cal G}(x))>
{\rm Corr}(f_1^{\cal D}(x),f_2^{\cal D}(x))$. Then, it is not difficult to see that
$$
D_{12}=\left(p_{12}^2\l_{12}+r_1p_{11}^2\l_{11} \right)
       \left(p_{21}^2\l_{12}+r_2 p_{22}^2\l_{22}\right)
      -\left(p_{12}^2\l_{12}+p_{11}^2\l_{11}\right)
       \left(p_{21}^2\l_{12}+p_{22}^2\l_{22}\right)
$$
with $r_k=(2-\l_{12})/(2-\l_{kk})$, $k=1,2$. We have the following cases:
\begin{enumerate}
\item
$\l_{12}>\max\{\l_{11},\l_{22}\}\,\Leftrightarrow\,r_1<1, r_2<1\,\Leftrightarrow\,
{\rm Corr}(f_1^{\cal G}(x),f_2^{\cal G}(x))>{\rm Corr}(f_1^{\cal D}(x),f_2^{\cal D}(x))$.
\item
$\l_{12}<\min\{\l_{11},\l_{22}\}\,\Leftrightarrow\,r_1>1, r_2>1\,\Leftrightarrow\,
{\rm Corr}(f_1^{\cal G}(x),f_2^{\cal G}(x))<{\rm Corr}(f_1^{\cal D}(x),f_2^{\cal D}(x))$.
\item
$\l_{12}=\l_{11}=\l_{22}\,\Leftrightarrow\,r_1=r_2=1\,\Leftrightarrow\,
{\rm Corr}(f_1^{\cal G}(x),f_2^{\cal G}(x))={\rm Corr}(f_1^{\cal D}(x),f_2^{\cal D}(x))$.
\end{enumerate}


\vspace{0.2in} \noindent{\bf 3. The PDGSBP Gibbs sampler.} In this section we will describe the PDGSBP Gibbs sampler for estimating the model. 
The details for the sampling algorithm of the PDDP model can be found in Hatjispyros et al. (2011, 2016A). 
At each iteration we will sample the variables,
\begin{align}
\nonumber
   & \theta_{jlk}, 1\le j \le l \le m,\,1\le k \leq N^*,\\
\nonumber
   & d_{ji},N_{ji},\delta_{ji}, 1\le j \le m,\,1\le i \le n_j,\\
\nonumber
   & p_{jl}, 1\leq j \le m, 1\le l \le m,
\end{align}
with $N^*=\max_{j,i}N_{ji}$ being almost surely finite. 

\medskip\noindent {\bf 1.} 
For the locations of the random measures for $k=1,\ldots,d^*$ where $d^*=\max_{j,i}d_{ji}$, it is that
$$
f(\theta_{jlk}|\cdots) \propto f(\theta_{jlk})
\begin{dcases}
       \prod_{i=1}^{n_j}K(x_{ji}|\theta_{jlk})^{{\cal I}(\d_{ji}=\vec{e}_l,\,d_{ji}=k)}
       \prod_{i=1}^{n_l}K(x_{li}|\theta_{jlk})^{{\cal I}(\d_{li}=\vec{e}_j,\,d_{li}=k)} & \,\,\,l>j\,,\\
       \prod_{i=1}^{n_j}K(x_{ji}|\theta_{jjk})^{{\cal I}(\d_{ji}=\vec{e}_j,\,d_{ji}=k)} & \,\,\,l=j\,.
\end{dcases}       
$$
If $N^*>d^*$ we sample additional locations $\t_{jl,d^*+1},\ldots,\t_{jl,N^*}$ independently from the prior. 

\medskip\noindent {\bf 2}. 
Here we sample the allocation variables $d_{ji}$ and the mixture 
component indicator variables $\delta_{ji}$ as a block. For $j=1,\ldots,m$ and $i=1,\ldots,n_{j}$, we have
$$
{\rm P}(d_{ji}=k,\d_{ji}=\vec{e}_l\,|N_{ji}=r,\cdots)\,\propto\, p_{jl}\,K(x_{ji}|\theta_{jlk})\,{\cal I}(l\le m)\,{\cal I}(k\le r).
$$

\medskip\noindent {\bf 3.} 
The slice variables $N_{ji}$ have full conditional distributions given by
$$
{\rm P}(N_{ji} = r\,|\,\d_{ji}=\vec{e}_l,d_{ji}=l,\cdots)\propto(1-\lambda_{jl})^r\,{\cal I}(r\ge l),
$$
which are truncated geometric distributions over the set $\{l, l+1,\ldots\}$.

\medskip\noindent {\bf 4.} 
The full conditional for $j=1,\ldots,m$ for the selection probabilities ${\boldsymbol p}_j=(p_{j1},\ldots,p_{jm})$, 
under a Dirichlet prior $f({\boldsymbol p}_j\,|\,{\boldsymbol a}_j)\propto\prod_{l=1}^m p_{jl}^{a_{jl}-1}$, with hyperparameter
${\boldsymbol a}_j=(a_{j1},\dots,a_{jm})$, is Dirichlet
$$
f({\boldsymbol p}_j\,|\cdots)\,\propto\,\prod_{l=1}^m p_{jl}^{a_{jl}+\sum_{i=1}^{n_l}{\cal I}(\delta_{ji}\,=\,\vec{e}_l) - 1}.
$$

\medskip\noindent {\bf 5.} 
Here we update the geometric probabilities $(\l_{jl})$ of the GSB measures. For $1\le j\le l\le m$, it is that
$$
f(\l_{jl}|\cdots) \propto f(\l_{jl})
\begin{dcases}
       \prod_{i=1}^{n_j}\left\{\l_{jl}^2(1-\l_{jl})^{N_{ji}-1}\right\}^{{\cal I}(\d_{ji}=\vec{e}_l)}
       \prod_{i=1}^{n_l}\left\{\l_{jl}^2(1-\l_{jl})^{N_{li}-1}\right\}^{{\cal I}(\d_{li}=\vec{e}_j)} & \,\,\,l>j\\
       \prod_{i=1}^{n_j}\left\{\l_{jj}^2(1-\l_{jj})^{N_{ji}-1}\right\}^{{\cal I}(\d_{ji}=\vec{e}_j)} & \,\,\,l=j\,.
\end{dcases}       
$$
To complete the model, we assign priors to the geometric probabilities.
For a fair comparison of the execution time between the two models, we apply $\l_{jl}=(1+c_{jl})^{-1}$
transformed priors. So, by placing gamma priors $c_{jl}\sim{\cal G}(a_{jl},b_{jl})$ 
over the concentration masses $c_{jl}$ of the PDDP model, we have
\begin{equation}
\label{TGamma}
f(\l_{jl})={\cal TG}(\l_{jl}\,|\,a_{jl},b_{jl})\propto
          \l_{jl}^{-(a_{jl}+1)}e^{-b_{jl}/\l_{jl}}(1-\l_{jl})^{a_{jl}-1}\,{\cal I}(0<\l_{jl}<1).
\end{equation}
In the Appendix, we give the full conditionals for $\l_{jl}$'s, their corresponding embedded 
Gibbs sampling schemes, and the sampling algorithm for the concentration masses.  


\vspace{0.2in} \noindent{\bf 3.1 The complexity of the rPDDP and PDGSBP samplers.} The main difference between the two samplers in terms of execution time,
comes from the blocked sampling of the clustering and the mixture indicator variables 
$d_{ji}$ and $\d_{ji}$.

\medskip\noindent{\bf The rPDDP model:}
The state space 
of the variable $(d_{ji},\delta_{ji})$ conditionally on the slice variable $u_{ji}$ is
$
(d_{ji},\delta_{ji})(\Omega)=
\cup_{l=1}^{m} \left(A_{w_{jl}}(u_{ji}) \times \{\vec{e}_l\}\right),
$
where $A_{w_{jl}}(u_{ji})=\{r\in{\mathbb N}:u_{ji}<w_{jlr}\}$ is the a.s. finite slice set corresponding to the observation $x_{ji}$ (Walker, 2007).
At each iteration of the Gibbs sampler, we have $m(m+1)/2$ vectors of stick-breaking weights $\vec{w}_{jl}$, each of length $N_{jl}^*$; 
where $N_{jl}^{*}\sim 1 + {\rm Poisson}(-c_{jl}\log u_{jl}^{*})$ with $c_{jl}$ being the concentration parameter of the Dirichlet 
process $\mathbb{P}_{jl}$ and $u_{jl}^{*}$ being the minimum of the slice variables in densities $f_{j}$ and $f_{l}.$ 
Algorithm $1$ gives the blocked sampling procedure of the clustering and mixture indicator variables. An illustration of 
the effect of the slice variable $u_{ji}$ is given in Figure 1(a).

\begin{algorithm}[H]
\caption{:~rPDDP}
\label{djiDPM}
\begin{algorithmic}[1]
\Procedure{Sample $(d_{ji},\d_{ji})$}{}
\For{random densities $f_{j},\,\,\,j=1$ to $m$ } 
\For{each data point $x_{ji}\in f_{j}\,\,\,i=1$ to $n_{j}$ }
\For{each mixture component $K(x_{ji}|\theta_{jl}),\,\,\,l=1$ to $m$ }
\State Construct slice sets $A_{w_{jl}}(u_{ji})$
\EndFor
\State Sample $(d_{ji}=k,\d_{ji}=r|\cdots)\propto K(x_{ji}|\theta_{jrk})\,{\cal I}\left((k,r)\in \cup_{l=1}^{m}\left(A_{w_{jl}}(u_{ji})\times \{\vec{e}_l\}\right)\right)$
\EndFor
\EndFor
\EndProcedure
\end{algorithmic}
\end{algorithm}

Since the weights forming the stick-breaking representation are not in an ordered form, the construction of the slice sets 
in step 5 of Algorithm \ref{djiDPM} requires a complete search in the array where the weights are stored. This operation 
is done in ${\cal O}(N_{jl}^{*})$ time. For the sampling of the $d_{ji}$ and $\d_{ji}$ variables in step 6, the choice of 
their value is an element from the union $\cup_{l=1}^{m}\left(A_{w_{jl}}(u_{ji})\times \{\vec{e}_l\}\right).$ This means that
the rPDDP algorithm for each $j$, must create $m$ slice sets which require $N_{jl}^{*}$ comparisons each. The worst case scenario is that 
the sampled $(d_{ji},\d_{ji})$ is the last element of $\cup_{l=1}^{m}\left(A_{w_{jl}}(u_{ji})\times \{\vec{e}_l\}\right)$. 
Thus, the DP based procedure of sampling $(d_{ji},\d_{ji})$ is of order 
$$
{\cal O}\left( m^2 n_j N_{jl}^* \sum_{l=1}^m|A_{w_{jl}}(u_{ji})| \right) 
= {\cal O}\left(N_{jl}^*\sum_{l=1}^m|A_{w_{jl}}(u_{ji})|\right).
$$


\medskip\noindent{\bf The PDGSBP model:}
The state space 
of the variable $(d_{ji},\delta_{ji})$ conditionally on the slice variable $N_{ji}$ is
$
(d_{ji}, \d_{ji})(\Omega)=\cup_{l=1}^{m} \left({\cal S}_{ji} \times \{\vec{e}_l\}\right).
$
In the GSB case, the slice variable has a different r\^olee. It indicates at which random 
point the search for the appropriate $d_{ji}$ will stop. In Figure 1(b) we illustrate this argument.
In Algorithm \ref{djiGSB} the worst case scenario is that the sampled $(d_{ji},\d_{ji})$ will be the last element 
of $\cup_{l=1}^{m} \left({\cal S}_{ji} \times \{\vec{e}_l\}\right)$. 
Thus, the GSB based procedure of sampling $(d_{ji},\d_{ji})$ is of order  
$
\mathcal{O}\left(m^{2}n_{j}N_{jl}\right) = \mathcal{O}\left(N_{jl}\right).
$

\begin{algorithm}[H]
\caption{:~PDGSBP}
\label{djiGSB}
\begin{algorithmic}[1]
\Procedure{Sample $(d_{ji},\d_{ji})$}{}
\For{random densities $f_{j},\,\,\,j=1$ to $m$ } 
\For{each data point $x_{ji} \in f_{j}\,\,\,i=1$ to $n_{j}$ }
\For{each mixture component $K(x_{ji}|\theta_{jl}),\,\,\,l=1$ to $m$ }
\State Sample $(d_{ji}=k,\d_{ji}=r|\cdots)\propto K(x_{ji}|\theta_{jrk})\,{\cal I}(k\leq N_{ji})\,{\cal I}(r\leq m)$
\EndFor
\EndFor
\EndFor
\EndProcedure
\end{algorithmic}
\end{algorithm}

\begin{figure}[H]
	\centering
	\subfloat[Subfigure 1 list of figures text][
	 Stick-breaking weights for some $N_{jl}^*=20$. The red dashed line represents 
	 the slice variable $u_{ji}=0.05$. 
	 The algorithm must check all the $N_{jl}^*$ values to accept those that they satisfy $u_{ji}<w_{jlk}$.
	 After a complete search, the slice set is $A_{w_{jl}}(u_{ji}) = \{1,2,3,5,7,8\}$.
   ]{
	\includegraphics[width=0.45\textwidth]{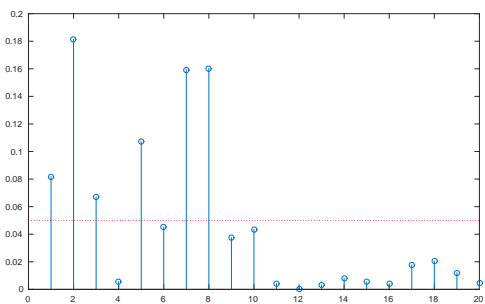}
	\label{fig:dpslice}}
	\qquad
	\subfloat[Subfigure 2 list of figures text][
	 Geometric stick-breaking weights for $N_{jl}^*=20$. The red dashed line represents the slice variable
	 $N_{ji}=6$. The slice set is simply ${\cal S}_{ji} = \{1,2,3,4,5,6\}$.]{
	\includegraphics[width=0.45\textwidth]{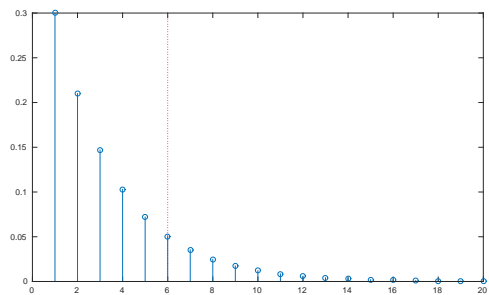}
	\label{fig:gsbslice}}
	\caption{A visualization of the effect of the $u_{ji}$ snd $N_{ji}$ slice variables are given 
	         in Figures 1(a) and 1(b) respectively.}
\label{fig:slicevars}
\end{figure}


\vspace{0.2in} \noindent{\bf 4. Illustrations.} In this section we illustrate the efficiency of the PDGSBP model. For the choice of a normal kernel 
(unless otherwise specified)
$K(x|\theta) = \mathcal{N}(x|\theta)$ where $\theta=(\mu,\tau^{-1})$ and $\tau=\sigma^{-2}$ is the precision. 
The prior over the means and precisions of the PDGSBP ($G_0$) and the rPDDP model ($P_0$) is the
independent normal-gamma measure, given by
$$
P_0(d\mu,d\tau)=G_0(d\mu,d\tau)=\,{\cal N}(\mu\,|\,\mu_0,\tau_0^{-1})
\,{\cal G}(\tau\,|\,\epsilon_1,\epsilon_2)\,d\mu d\tau.
$$
Attempting a noninformative prior specification (unless otherwise specified),
we took $\mu_0=0$ and $\tau_0=\epsilon_1=\epsilon_2=10^{-3}$.
For the concentration masses of the rPDDP model, a-priori, we set $c_{jl}\sim{\cal G}(a_{jl},b_{jl})$.
For an objective evaluation of the execution time, of the two algorithms under different 
scenarios, we choose a synchronized prior specification, namely, for the geometric probabilities, 
we set $\l_{jl}\sim{\cal TG}(a_{jl},b_{jl})$ -- the transformed gamma density given in equation (\ref{TGamma}). 
In the appendix B, we show that such prior specifications are valid for $a_{jl}>1$.
In all our numerical examples, we took $a_{jl}=b_{jl}=1.1$.
For our numerical experiments (unless otherwise specified), the hyperparameters $(\a_{jl})$ of the Dirichlet 
priors over the matrix of the selection probabilities $p=(p_{jl})$ has been set to $\a_{jl}=1$.

In all cases, we measure the similarity between probability distributions with the Hellinger distance.
So for example, ${\cal H}_{\cal G}(f,\hat{f})$ and ${\cal H}_{\cal D}(f,\hat{f})$, will denote the Hellinger 
distance between the true density $f$ and the predictive density $\hat{f}$ of the PDGSBP and rPDDP algorithms,
respectively. The Gibbs samplers run for $11\times 10^4$ iterations leaving 
the first $10^4$ samples as a burn-in period. 


\medskip\noindent{\bf 4.1 Time execution efficiency of the PDGSBP model.} 

\medskip\noindent{\bf Nested normal mixtures with a unimodal common and idiosyncratic part:} 
Here, we choose to include all pairwise and idiosyncratic
dependences in the form of unimodal equally weighted normal mixture components. The mixture
components are well separated with unit variance. We define each data model 
${\cal M}_m=\{f_j^{(m)}:1\le j\le m\}$ of dimension $m\in\{2,3,4\}$, based on a 
$4\times 10$ matrix $M=(M_{jk})$, with entries in the set $\{0,1\}$, having at most two ones in 
each column and exactly four ones in each row. When there is exactly one entry of one, the column defines 
an idiosyncratic part. The appearance of exactly two ones in a column defines a common component.
We let the matrix $M$ given by  
\[
M=
  \begin{bmatrix}
    1 & 1 & 1 & 1 & 0 & 0 & 0 & 0 & 0 & 0 \\
    0 & 0 & 1 & 0 & 1 & 0 & 0 & 1 & 1 & 0 \\
    0 & 1 & 0 & 0 & 0 & 1 & 0 & 1 & 0 & 1 \\
    1 & 0 & 0 & 0 & 0 & 0 & 1 & 0 & 1 & 1 \\
  \end{bmatrix},
\]
and for $m\in\{2,3,4\}$, we define 
$$
{\cal M}_m:\,f_j^{(m)}(x)\propto\sum_{k=5-m}^{2(m+1)}M_{jk}\,{\cal N}(x|10(k-6),1),\,\,1\le j\le m,
$$
We are taking independently samples of sizes $n_j^{(2)}=60$ from the $f_j^{(2)}$'s, 
$n_j^{(3)}=120$ from the $f_j^{(3)}$'s, and, $n_j^{(4)}=200$ from the $f_j^{(4)}$'s.
In all cases, the PDGSBP and the rPDDP density estimations are of the same quality.

In Figures 2(a)--(d) we give the histograms of the data sets for the specific case $m=4$, 
which are overladed with the kernel density estimations (KDE's) based on the predictive samples of the 
$f_j^{(4)}$'s coming from the PDGSBP (solid line) and the rPDDP (dashed line) models.
The differences between the two models are nearly indistinguishable. 
The Hellinger distances between the true and the estimated densities for the case $m=4$ are given in table 1.

In Table 2 we summarize the mean execution times (MET's) per $10^3$ iterations in seconds. 
The PDGSBP sampler is about three times faster than the rPDDP sampler. The corresponding 
MET ratios for $m=2,3$ and 4 are $2.96, 3.04$ and 3.37 respectively.
We can see that the PDGSBP Gibbs sampler gives slightly faster execution times with increasing
$m$. This will become more clear in our next simulated data example, where the average sample 
size per mode is being kept constant. 

\begin{figure}[H]
\includegraphics[width=1\textwidth]{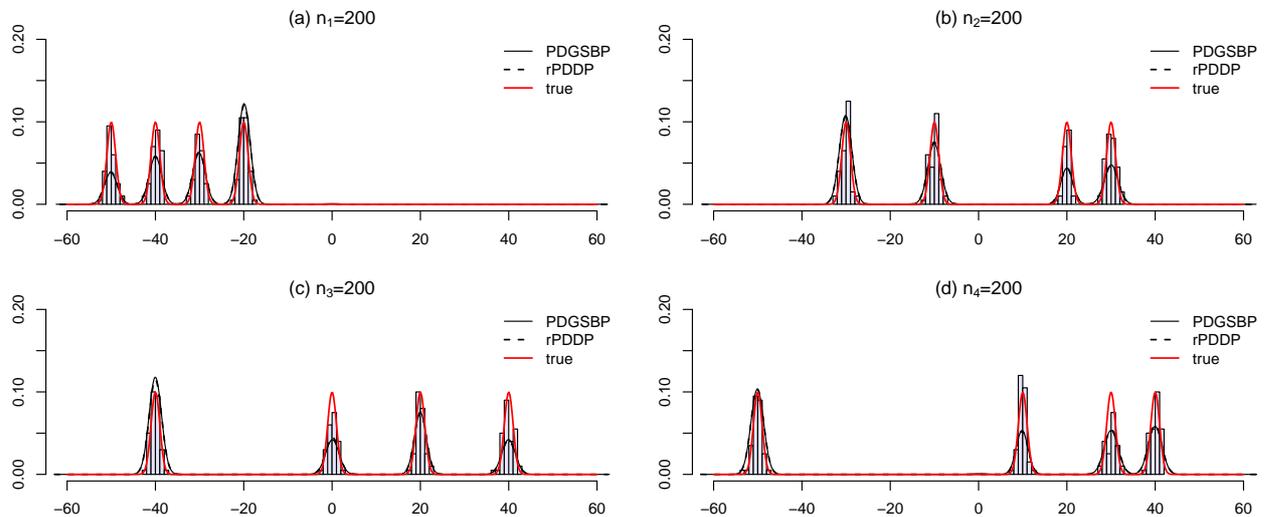}
\centering
\caption{Histograms of data sets coming for the case $m=4$. The superimposed KDE's are
based on the predictive samples obtained from the PDGSBP and the rPDDP models.}
\end{figure}

\begin{table}[H]
\begin{tabular}{ccc} 
\hline
$i$ & ${\cal H}_{\cal{G}}(f_i^{(4)},\hat{f}_i^{(4)})$ & ${\cal H}_{\cal{D}}(f_i^{(4)},\hat{f}_i^{(4)})$\\
\hline\hline
$1$ & $0.17$ & $0.17$ \\ 
$2$ & $0.19$ & $0.18$ \\ 
$3$ & $0.22$ & $0.22$ \\
$4$ & $0.20$ & $0.20$ \\
\hline
\end{tabular}
\caption{Hellinger distances for the case $m=4$.}
\end{table}

\begin{table}[H]
\begin{tabular}{cllr}
\hline
   $m$   &    Model    & Sample size   &   MET        \\
\hline\hline
    2    &    PDGSBP   & $n_j^{(2)}=60$      &   0.57       \\
         &    rPDDP    &               &   1.68       \\   
\hline          
    3    &    PDGSBP   & $n_j^{(3)}=120$     &   2.16       \\
         &    rPDDP    &               &   6.57       \\
\hline         
    4    &    PDGSBP   & $n_j^{(4)}=200$     &   5.30       \\
         &    rPDDP    &               &   17.87      \\    
\hline
\end{tabular}
\caption{Mean execution times in seconds per $10^3$ iterations.}
\end{table}

\medskip\noindent{\bf Sparse $m$--scalable data set models:} 
In this example, we attempt to create $m$-scalable normal mixture data sets of the lowest possible sample size. 
To this respect, we sample independently $m$ groups of data sets from the densities
$$
f_j^{(m)}(x)\,\propto\,{\cal N}(x|(j-1)\xi,1)\,{\cal I}(1\le j<m)+
\sum_{k=1}^{m-1}{\cal N}(x|(k-1)\,\xi,1)\,{\cal I}(j=m),
$$
with sample sizes 
$
n_j^{(m)}=n\{{\cal I}(1\le j<m)+(m-1)\,{\cal I}(j=m)\}.
$
We have chosen $\xi=10$ and an average sample size per mode of $n=20$, for $m\in\{2,\ldots,10\}$.

In Figure 3 we depict the average execution times as functions of the dimension $m$.
We can see how fast the two MET-curves diverge with increasing $m$. 
In Figure 4(a)--(j), for the specific case $m=10$, we give the histograms of the data sets,  
overladed with the KDE's based on the predictive samples of the $f_j^{(10)}$'s coming from the 
PDGSBP (solid line) and the rPDDP (dashed line) models. 
We can see that the PDGSBP and the rPDDP density estimations are of the same quality.

The Hellinger distances between the true and the estimated 
densities for the specific case $m=10$ are given in Table 3. 
The large values of the Hellinger distances
${\cal H}_{\cal{G}}(f_{10}^{(10)},\hat{f}_{10}^{(10)})\approx 
{\cal H}_{\cal{D}}(f_{10}^{(10)},\hat{f}_{10}^{(10)})\approx 0.22$,
are caused by the enlargement of the variances of the underrepresented modes due to the small sample size.

\begin{figure}[H]
\includegraphics[width=0.6\textwidth]{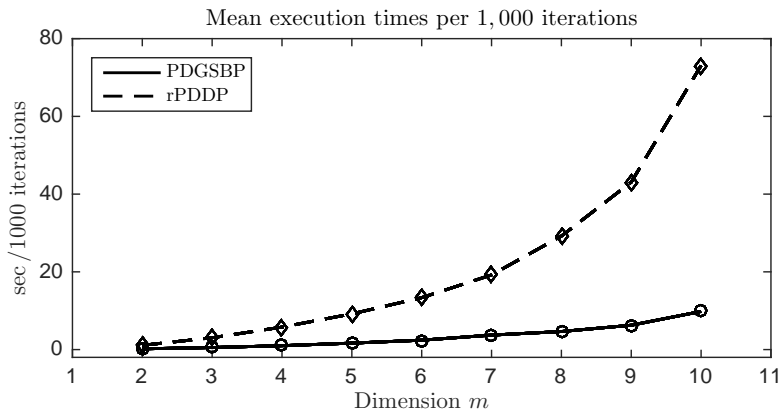}
\centering
\caption{Mean execution times for the two models, based on the sparse $m$-scalable data sets.}
\end{figure}

\begin{figure}[H]
\includegraphics[width=1\textwidth]{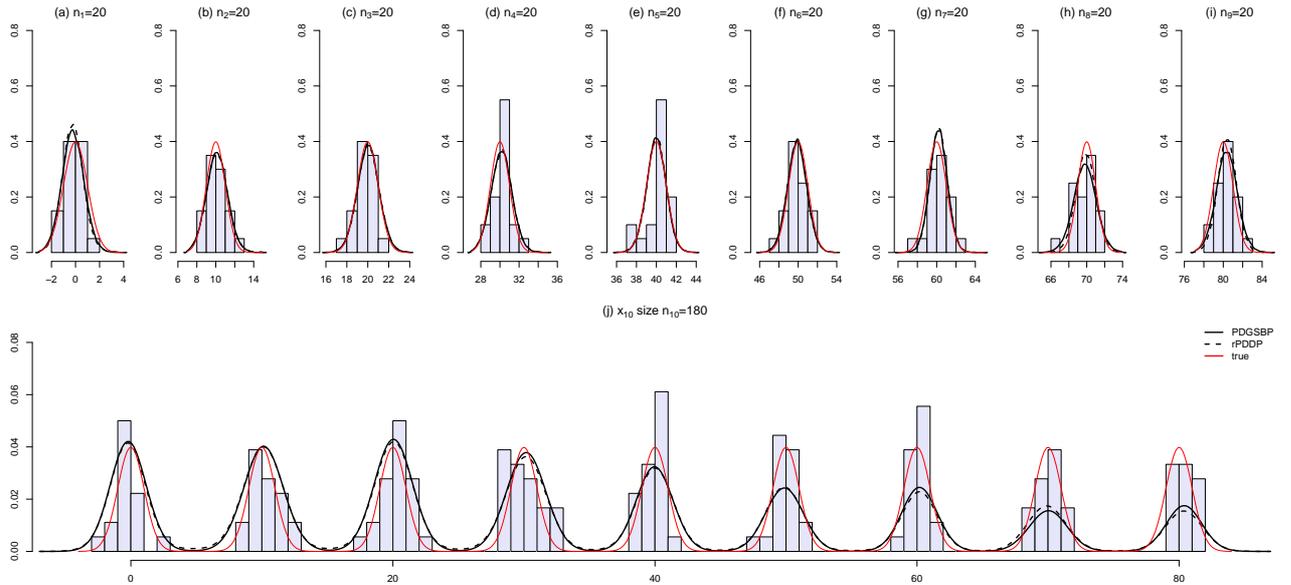}
\centering
\caption{Histograms of sparse $m$-scalable data sets for the case $m=10$. 
         The superimposed KDE's are based on the predictive samples of the 
         PDGSBP and the rPDDP models.}
\end{figure}

\begin{table}[H]
\begin{tabular}{ccccccccccc} 
\hline
$ i$ & $1$ & $2$ & $3$ & $4$ & $5$ & $6$ & $7$ & $8$ & $9$ & $10$  \\
\hline\hline
${\cal H}_{\cal{G}}(f_i^{(10)},\hat{f}_i^{(10)})$ & $0.08$ & $0.10$ & $0.09$ & $0.14$ & $0.14$ & $0.13$ & $0.14$ & $0.09$ & $0.11$ & $0.22$\\ 
${\cal H}_{\cal{D}}(f_i^{(10)},\hat{f}_i^{(10)})$ & $0.09$ & $0.11$ & $0.10$ & $0.15$ & $0.12$ & $0.10$ & $0.14$ & $0.09$ & $0.09$ & $0.22$\\ 
\hline
\end{tabular}
\caption{Hellinger distances between true and estimated densities  
        for the case $m=10$ of the sparse scalable data example.}
\end{table}


\medskip\noindent{\bf 4.2 Normal and gamma mixture models that are not well separated.} 

\medskip\noindent
{\bf The normal mixture example:} 
We will first consider a normal model for $m=2$, first appeared in Lijoi et. al (2014B).
The data models for $f_1$ and $f_2$ are 7-mixtures. Their common part is a 4-mixture that is weighted differently
between the two mixtures. More specifically, we sample two data sets of sample size $n_1=n_2=200$,  independently from 
$$
(f_1,f_2)=\left({1\over 2}\, g_{11} + {1\over 2}\, g_{12},\,\,{4\over 7}\, g_{21} + {3\over 7}\,g_{22}\right), 
$$
with
\begin{align}
g_{11} &= \frac{2}{7}{\cal N}(-8,0.25^2) + \frac{3}{7}{\cal N}(1,0.5^2) + \frac{2}{7}{\cal N}(10,1) \nonumber\\
g_{12} &= \frac{1}{7}{\cal N}(-10,0.5^2) + \frac{3}{7}{\cal N}(-3,0.75^2) + \frac{1}{7}{\cal N}(3,0.25^2) + \frac{2}{7}{\cal N}(7, 0.25^2)\nonumber\\
g_{21} &= \frac{2}{8}{\cal N}(-10,0.5^2) + \frac{3}{8}{\cal N}(-3,0.75^2) + \frac{2}{8}{\cal N}(3,0.25^2) + \frac{1}{8}{\cal N}(7, 0.25^2)\nonumber\\
g_{22} &= \frac{1}{3}{\cal N}(-6,0.5^2) + \frac{1}{3}{\cal N}(-1,0.25^2) + \frac{1}{3}{\cal N}(5,0.5^2).\nonumber 
\end{align}
For this case, a-priori we took $(\mu_0,\tau_0,\epsilon_1,\epsilon_2)=(0,10^{-3},1,10^{-2})$. 

In Figure 5(a)--(b) we give the histograms of the data sets, 
with the predictive densities of the PDGSBP and rPDDP models 
superimposed in black solid and black dashed curves, respectively.
We can see that the PDGSBP and the rPDDP density estimations are of the same quality.
In Table 4, we give the Hellinger distance between the true and the estimated densities

\begin{figure}[H]
\centering
\includegraphics[width=0.9\textwidth]{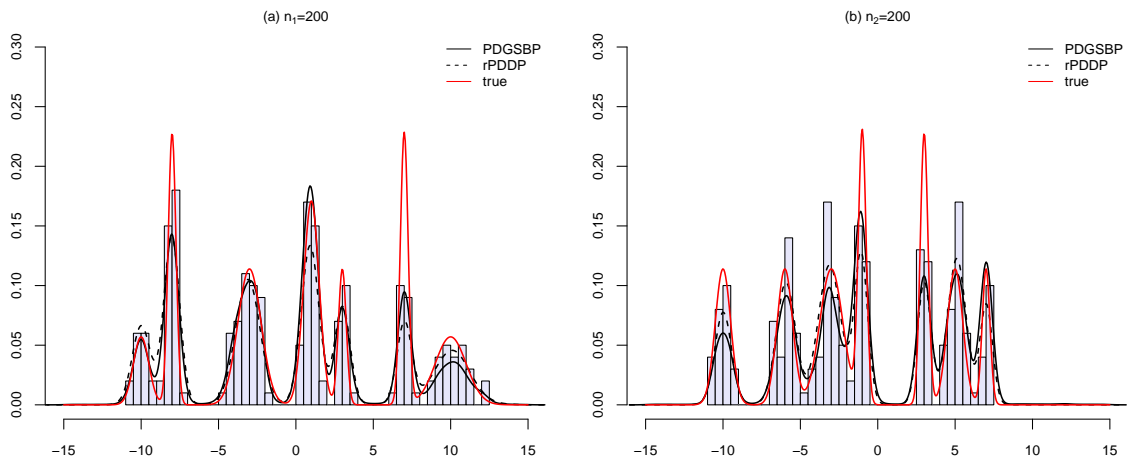}
\caption{Density estimations of the 7-mixtures data sets, under the PDGSBP and the rPDDP models. 
The true densities have been superimposed in red.} 
\end{figure}

\begin{table}[H]
\begin{tabular}{ccc} 
\hline
$ i$ & ${\cal H}_{\cal G}(f_i,\hat{f}_i)$ & ${\cal H}_{\cal D}(f_i,\hat{f}_i)$  \\
\hline\hline
$1$ & $0.19$ & $0.18$ \\ 
$2$ & $0.18$ & $0.15$ \\ 
\hline
\end{tabular}
\caption{Hellinger distance between the true and the estimated densities.}
\end{table}

\medskip\noindent
{\bf The gamma mixture example:} 
In this example we took $m=2$. The data models for $f_1$ and $f_2$ are gamma 4-mixtures. 
The common part is a gamma 2-mixture, weighted identically among the two mixtures. 
More specifically, we sample two data sets of sample size $n_1=n_2=160$,  independently from 
$$
(f_1,f_2)=\left({2\over  5}\, g_{11} + {3\over  5}\,g_{12},\,\,
                {7\over 10}\, g_{12} + {3\over 10}\,g_{22}\right), 
$$
with
\begin{align}
g_{11} &=  \frac{2}{ 3}{\cal G}(  2,1.1) + \frac{1}{ 3}{\cal G}( 80,2)  \nonumber\\
g_{12} &=  \frac{8}{14}{\cal G}( 10,0.9) + \frac{6}{14}{\cal G}(200,8.1)\nonumber\\
g_{22} & = \frac{2}{ 3}{\cal G}(105,3)   + \frac{1}{ 3}{\cal G}(500,10),\nonumber
\end{align}
Because we want to estimate the density of non negative observations, we find it more 
appropriate to take the kernel to be a log-normal distribution (Hatjispyros et al. 2016B).
That is $K(x|\theta) = \mathcal{LN}(x|\theta)$ with $\theta=(\mu,\sigma^2)$, 
is the log-normal density with mean $\exp(\mu+\sigma^{2}/2)$. For this case, a-priori we set 
$$
(\mu_0,\tau_0,\epsilon_1,\epsilon_2)=(\bar{S},0.5,2,0.01),\quad
\bar{S}={1\over n_1+n_2}\left(\sum_{j=1}^{n_1}\log x_{1j}+\sum_{j=1}^{n_2}\log x_{2j}\right).
$$
In Figure 6(a)-(b), we display the KDE's based on the predictive samples of the two models.
We can see that the PDGSBP and the rPDDP density estimations are of the same quality.
In Table 5, we give the Hellinger distances. 

\begin{figure}[H]
\centering
\includegraphics[width=0.9\textwidth]{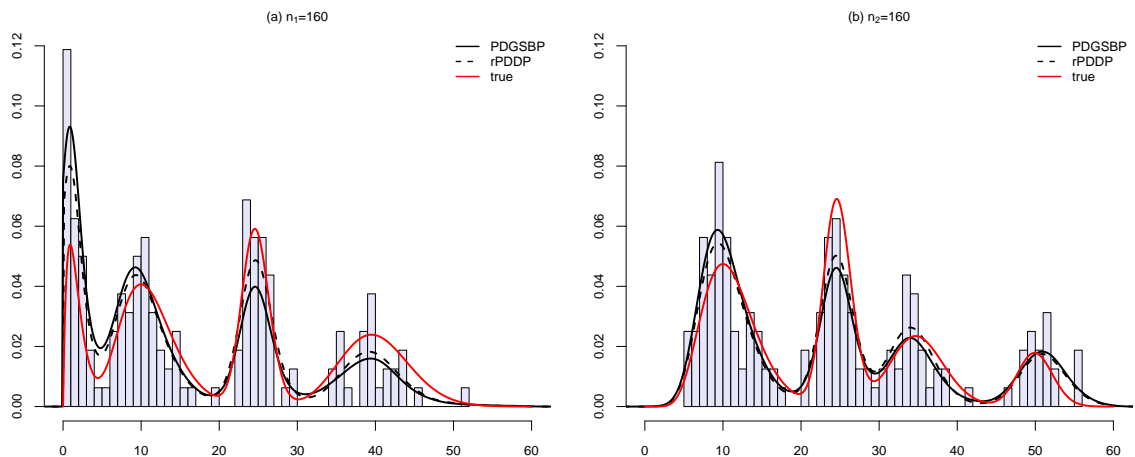}
\caption{The KDE's are based on the predictive sample of the PDGSBP model (solid curve in black)
         and the predictive sample of the rPDDP model (dashed curve in black).} 
\end{figure}

\begin{table}[H]
\begin{tabular}{ccc} 
\hline
$ i$ & ${\cal H}_{\cal G}(f_i,\hat{f}_i)$ & ${\cal H}_{\cal D}(f_i,\hat{f}_i)$  \\
\hline\hline
$1$ & $0.13$ & $0.11$ \\ 
$2$ & $0.19$ & $0.18$ \\ 
\hline
\end{tabular}
\caption{Hellinger distances for the gamma mixture data model.}
\end{table}

Because the common part is equally weighted 
among $f_1$ and $f_2$, it makes sense to display the estimations of the selection probability
matrices under the two models
$$
\E_{\cal G}(p\,|\,(x_{ji})) = \begin{pmatrix} 0.42 & 0.58\\
											                        0.64 & 0.36
											        \end{pmatrix},\quad
\E_{\cal D}(p\,|\,(x_{ji})) = \begin{pmatrix} 0.42 & 0.58\\
											                        0.69 & 0.31
											        \end{pmatrix},\quad
                 p_{\rm true}=\begin{pmatrix} 0.4 & 0.6\\
					                                    0.7 & 0.3
					                    \end{pmatrix}.											        											 
$$


\medskip\noindent{\bf 4.3 Borrowing of strength of the PDGSBP model.} In this example we consider three populations $\{D_j^{(s)}:j=1,2,3\}$, 
under three different scenarios $s\in\{1,2,3\}$. The sample sizes are always the same, namely, 
$n_1=200$, $n_2=50$ and $n_3=200$ -- the second population is sampled only once.
The three data sets $D_1^{(s)}$, $D_2^{(s)}$ and $D_3^{(s)}$, are sampled independently 
from the normal mixtures
$$
(f_{1}^{(s)},f_{2}^{(s)},f_{3}^{(s)})=
\left((1-q^{(s)})f+q^{(s)}g_1,\,\,f,\,\,(1-q^{(s)})f+q^{(s)}g_2\right),
$$
where 
\begin{align}
f\,\, & =\frac{3}{10}{\cal N}(-10,1) + \frac{2}{10}{\cal N}(-6,1) 
      +\frac{2}{10}{\cal N}(6,1) + \frac{3}{10}{\cal N}(10,1)\nonumber\\
g_1 & =\,\frac{1}{2}{\cal N}(-4,1) + \frac{1}{2}{\cal N}(4,1)\nonumber\\
g_2 & =\,\frac{1}{2}{\cal N}(-12,1) + \frac{1}{2}{\cal N}(12,1).\nonumber
\end{align}

More specifically, the three scenarios are: 
\begin{enumerate}

\item 
For $s=1$, we set, $q^{(1)}=0$. This is the case where the three populations are coming from 
the same 4--mixture $f$.  We depict the density estimations under the first scenario in Figures 7(a)--(c).
This is the case where the small data set, benefits the most in terms of borrowing of strength.

\item 
For $s=2$, we set, $q^{(2)}=1/2$. The 2-mixtures $g_1$ and $g_2$ are the the idiosyncratic
parts of the 6-mixtures $f_1^{(2)}$ and $f_3^{(2)}$, respectively. 
The density estimations under the second scenario are given in Figures 7(d)--(f).
In this case, the strength of borrowing 
between the small data set and the two large data sets weakens.

\item 
For $s=3$ we set $q^{(3)}=1$. In this case the three populations have no common
parts. The density estimations are given in Figures 7(g)--(i).
This is the worst case scenario, where there is no borrowing
of strength between the small and the two large data sets.
\end{enumerate}

The Hellinger distances between the true and the estimated 
densities, for the three scenarios, are given in table 6.
In the second column of the Table we can see how the Hellinger distance of the 
estimation $\hat{f}_2^{(s)}$ and the true density $f_2^{(s)}$ increases as the borrowing of strength 
weakens, it is that
$
{\cal H}_{\cal G}(f_2^{(1)},\hat{f}_2^{(1)})<{\cal H}_{\cal G}(f_2^{(2)},\hat{f}_2^{(2)})
<{\cal H}_{\cal G}(f_2^{(3)},\hat{f}_2^{(3)}).
$

\begin{figure}[H]
\centering
\includegraphics[width=0.9\textwidth]{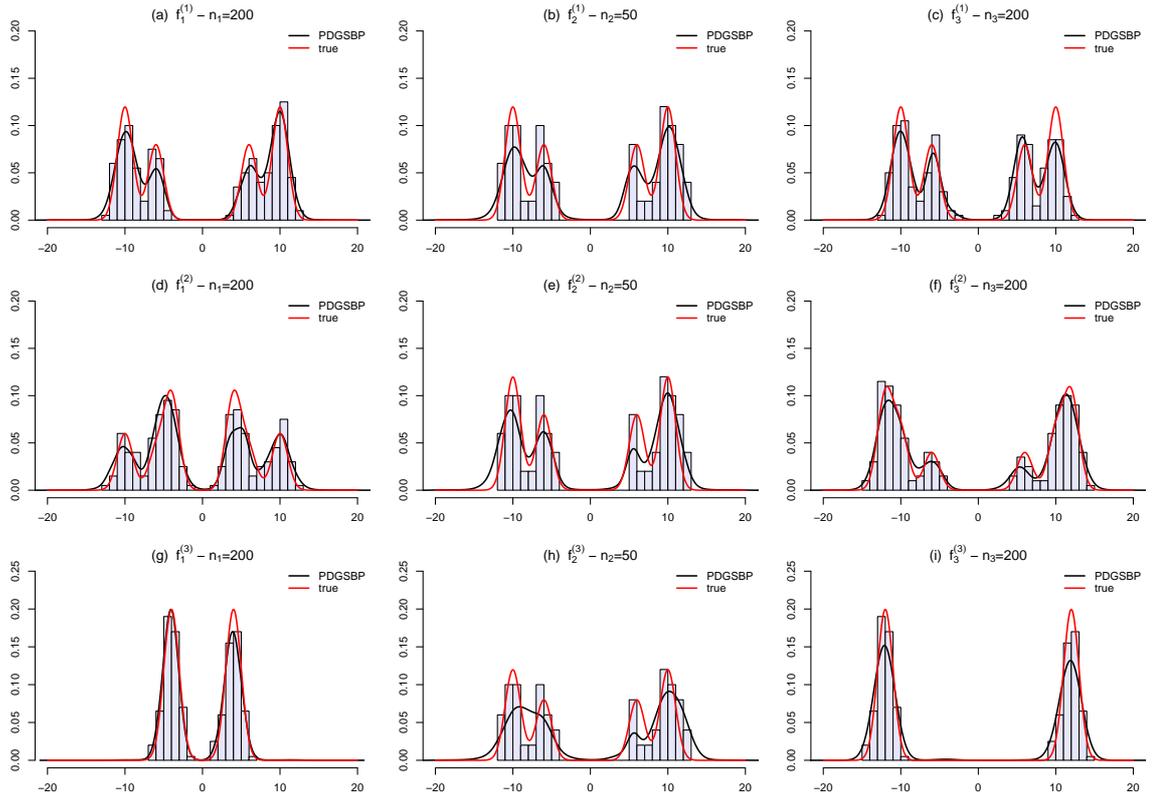}
\caption{Density estimation with the PDGSBP model (curves in black) under the three 
         different scenarios. The true density has been superimposed in red.} 
\end{figure}

\begin{table}[H]
\begin{tabular}{cccc} 
\hline
$s$ & ${\cal H}_{\cal{G}}(f_1^{(s)},\hat{f}_1^{(s)})$ & ${\cal H}_{\cal{G}}(f_2^{(s)},\hat{f}_2^{(s)})$ & ${\cal H}_{\cal{G}}(f_3^{(s)},\hat{f}_3^{(s)})$  \\
\hline\hline
$1$ & $0.14$ & $0.19$ & $0.13$\\ 
$2$ & $0.15$ & $0.22$ & $0.15$\\ 
$3$ & $0.12$ & $0.26$ & $0.12$\\ 
\hline
\end{tabular}
\caption{Hellinger distances between the true and the estimated densities for the three scenario example.}
\end{table}


\medskip\noindent{\bf 4.4 Real data example.}  The data set is to be found at \texttt{http://lib.stat.cmu.edu/datasets/pbcseq} and involves data 
from 310 individuals. We take the observation as SGOT (serum glutamic-oxaloacetic transaminase) 
level, just prior to liver transplant or death or the last observation recorded, under three conditions on the individual
\begin{enumerate}
\item The individual is dead without transplantation.
\item The individual had a transplant.
\item The individual is alive without transplantation.
\end{enumerate}
We normalize the means of all three data sets to zero. Since it is reasonable 
to assume the densities for the observations are similar for the three categories 
(especially for the last two), we adopt the models proposed in this paper with $m = 3$. 
The number of transplanted individuals is small (sample size of 28) so it is reasonable 
to borrow strength for this density from the other two. In this example, we set the hyperparameters of 
the Dirichlet priors for the selection probabilities to 
$$
\a_{jl}= \begin{cases} 10, & \mbox{if } j=l=1 \mbox{ or } j=l=3\\ 1, & \mbox{ otherwise.} \end{cases}
$$

\begin{enumerate}
\item In Figure 8(a)--(c) we provide histograms of the real data sets and superimpose the KDE's
      based on the predictive samples of the PDDP and PDGSBP samplers. The two models give nearly 
      identical density estimations.
      
\item The estimated a-posteriori selection probabilities are given below 
$$
\E_{\cal G}(p\,|\,(x_{ji})) = \begin{pmatrix} 
                       0.61  & 0.23 & 0.16\\
                       0.34  & 0.10 & 0.56\\
                       0.08  & 0.12&  0.80
                       \end{pmatrix},\quad
\E_{\cal D}(p\,|\,(x_{ji}))  = \begin{pmatrix} 
                       0.67  & 0.16 & 0.17\\
                       0.29  & 0.15 & 0.56 \\ 
                       0.10  & 0.12 & 0.78
                       \end{pmatrix}.				          
$$
\end{enumerate}
By comparing the second rows of the selection matrices, we conclude that the strength of borrowing is slightly larger in the case of PDGSBP model . 

\begin{figure}[H]
\includegraphics[width=1\textwidth]{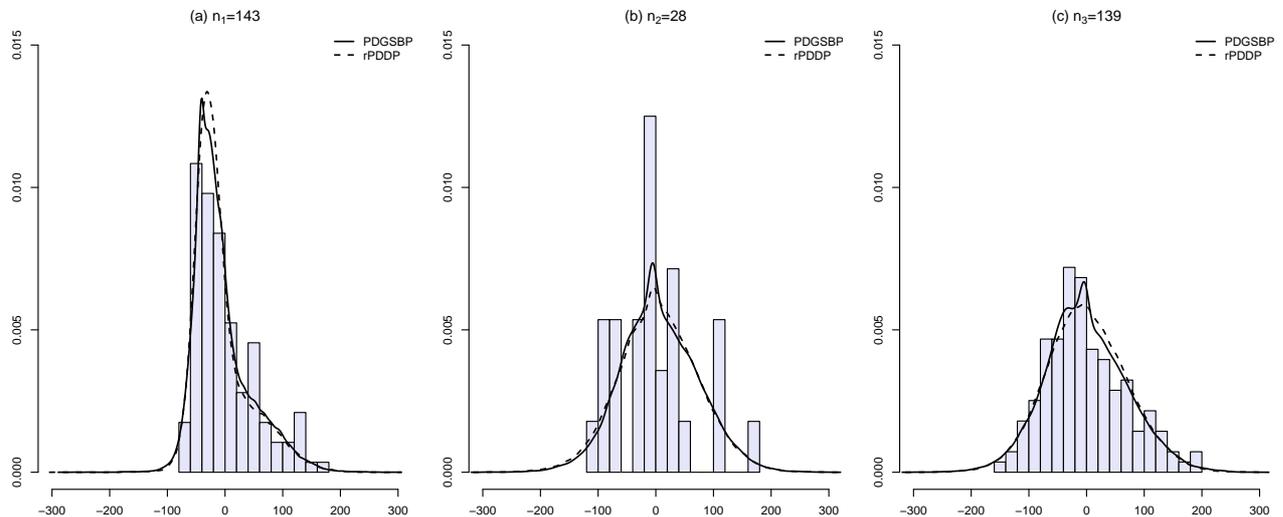}
\centering
\caption{Histograms of the real data sets with superimposed KDE curves
         based on the predictive samples of the PDGSBP and rPDDP models.}
\end{figure}


\vspace{0.2in} \noindent{\bf 5. Discussion.} In this paper we have generalized the GSB process to a multidimensional dependent 
stochastic process which can be used as a Bayesian nonparametric prior for density estimation in the case 
of partially exchangeable data sets.
The resulting Gibbs sampler is as accurate as its DP based counterpart, yet faster and far less complicated.
The main reason for this is that the GSB sampled value of the allocation variable $d_{ji}$ will be an 
element of the sequential slice set 
${\cal S}_{ji}=\{1,\ldots,N_{ji}\}$. Thus, there is no need to search the arrays of the weights; 
we know the state space of the clustering variables in advance. On the other hand, the sampling of $d_{ji}$ in the DP based 
algorithm will always have one more step; the creation of the slice sets.

For an objective comparison of the execution times of the two models,
we have run the two samplers in an a-priori synchronized mode. 
This, involves the placing of ${\cal G}(a_{jl},b_{jl})$ priors over the DP $c_{jl}$
concentration masses, leading to a more efficient version of the PDDP model introduced in 
Hatjispyros et al. (2011, 2016A).

We have show that when the PDGSBP and PDDP models are synchronized, i.e. their parameters satisfy
$\l_{ji}=(1+c_{ji})^{-1}$, the correlation between the models can be controlled by
imposing further restrictions among the $\l_{ji}$ parameters.

Finally, an interesting research path would be the generalization of the pairwise dependent
$\Q_j$ measures to include all possible interactions, in the sense that
$$
\Q_j(\,\cdot\,)=p_j\,\G_j(\,\cdot\,)+\sum_{l=2}^m\sum_{\eta\,\in\,{\cal C}_{j,l,m}}p_{j,\eta}\,\G_{\eta_{(j)}}(\,\cdot\,)\quad
{\rm with}\quad
p_j+\sum_{l=2}^m\sum_{\eta\,\in\,{\cal C}_{j,l,m}}p_{j,\eta}=1,
$$
where the $\G_j$ and the $\G_{\eta_{(j)}}$'s are independent GSB processes,
${\cal C}_{j,l,m}=\{(k_1,\ldots,k_{l-1}):1\le k_1<\cdots<k_{l-1}\le m, k_r\neq j, 1\le r\le m-1\}$
and $\eta_{(j)}$ is the ordered vector of the elements of the vector $\eta$ and $\{j\}$.
Now the $f_j$ densities will be a mixture of $2^{m-1}$ GSB mixtures, and
the total number of the independent GSB processes needed to model $(f_1,\ldots,f_m)$
will be $2^m-1$.


\vspace{0.2in} \noindent{\bf Appendix A}

\noindent{\bf Proof of Proposition 1.}
Starting from the $N_{ji}$-augmented random densities we have
{\footnotesize
\begin{eqnarray}
f_j(x_{ji},N_{ji}=r) & = & \sum_{l=1}^{m}f_{j}(x_{ji},N_{ji}=r,\delta_{ji}=l) = \sum_{l=1}^{m}p_{jl}\,f_{j}(x_{ji},N_{ji}=r|\delta_{ji} = l) \nonumber\\
& = & \sum_{l=1}^{m}p_{jl}\sum_{k=1}^{\infty}f_{j}(x_{ji},N_{ji}=r,d_{ji}=k|\delta_{ji}=l) \nonumber\\
& = & \sum_{l=1}^{m}p_{jl}f_{j}(N_{ji}=r|\delta_{ji}=l)\sum_{k=1}^{\infty}f_j(d_{ji}=k|N_{ji}=r)f_{j}(x_{ji}|d_{ji}=k,\delta_{ji}=l).\nonumber
\end{eqnarray}
}%
Because $f_j(N_{ji}=r|\d_{ji}=l)=f_N(r|\lambda_{jl})$ and
$f_j(x_{ji}|d_{ji}=k,\delta_{ji}=l)=K(x_{ji}|\t_{jlk})$, the last equation gives
\begin{align}
 \nonumber
   & f_j(x_{ji},N_{ji}=r)  =  \sum_{l=1}^m p_{jl} f_N(r|\lambda_{jl})\sum_{k=1}^\infty {1\over r}{\cal I}(k\le r)K(x_{ji}|\t_{jlk})\\
 \nonumber
   & ~~~~~~~~~~~~~~~~~~~\, =  \frac{1}{r} \sum_{l=1}^{m}p_{jl}f_{N}(r|\lambda_{jl})\sum_{k=1}^{r}\,K(x_{ji}|\theta_{jlk}).
\end{align}
Augmenting further with the variables $d_{ji}$ and $\d_{ji}$ yields
$$
  f_j(x_{ji},N_{ji}=r,d_{ji}=k,\d_{ji}=l)={1\over r}\,p_{jl}\,f_N(r|\lambda_{jl})\,{\cal I}(k\le r)\,K(x_{ji}|\t_{jlk}).
$$
Because P$(\d_{ji}=l)=p_{jl}$, the last equation leads to equation (\ref{proposition12}) and the proposition follows. \hfill$\square$

\medskip\noindent{\bf Proof of Proposition 2.}
Marginalizing the joint of $x_{ji}$ and $N_{ji}$ with respect to $x_{ji}$ we obtain
$$
f_j(N_{ji}=r)=\sum_{l=1}^m p_{jl}f_N(r|\lambda_{jl}).
$$
Then dividing equation (\ref{proposition11}) with the probability that $N_{ji}$ equals $r$ we obtain equation (\ref{proposition21}). \hfill $\square$

\medskip\noindent{\bf Proof of Lemma 1.}
Because $g_\G(x)=\l\sum_{j=1}^\infty(1-\l)^{j-1}K(x|\t_j)$, we have
\begin{align}
\nonumber
 & \E\left\{g_\G(x)^2\right\}=\l^2\,\E\left\{\left(\sum_{j=1}^{\infty}(1-\l)^{j-1}K(x|\theta_j)\right)^2\right\}\\
\nonumber
 & =\l^2\left\{\sum_{j=1}^\infty(1-\l)^{2j-2}\,\E\left[K(x|\t_j)^2\right]+2\sum_{k=2}^{\infty}\sum_{j=1}^{k-1}(1-\l)^{j+k-2}\,\E[K(x|\t_j)K(x|\t_k)]\right\}\\
\nonumber 
 & =\l^2\left\{\sum_{j=1}^\infty(1-\l)^{2j-2}\E\left[K(x|\t)^2\right]+2\,\sum_{k=2}^{\infty}\sum_{j=1}^{k-1}(1-\l)^{j+k-2}\E[K(x|\t)]^2\right\}\\
\nonumber 
 & =\l^2\left\{{1\over \l(2-\l)}\E\left[K(x|\t)^2\right]+2\,{1-\l\over \l^2(2-\l)}\E[K(x|\t)]^2\right\}, 
\end{align}
which gives the desired result.\hfill$\square$

\medskip\noindent{\bf Proof of Proposition 3.}
The random densities $f_i(x)=\sum_{l=1}^m p_{il}\,g_{il}(x)$ and $f_j(x)=\sum_{l=1}^m p_{jl}\,g_{jl}(x)$ 
depend to each other through the random measure $\G_{ji}$, therefore 
\begin{equation}
\label{fifj1}
\E[f_i(x)f_j(x)]= \E[\,\E(f_i(x)f_j(x)|\G_{ji})\,]=\E\{\,\E[f_i(x)|\G_{ji}]\,\E[f_j(x)|\G_{ji}]\,\},
\end{equation}
and
\begin{align}
\nonumber
   &   \E[f_j(x)|\G_{ji}]=\sum_{l\neq i}p_{jl}\,\E[g_{jl}(x)]+p_{ji}g_{ji}(x)
                         =(1-p_{ji})\,\E[K(x|\t)]+p_{ji}g_{ji}(x)\\
\nonumber
   &   \E[f_i(x)|\G_{ji}]=\sum_{l\neq j}p_{il}\,\E[g_{il}(x)]+p_{ij}g_{ji}(x)
                         =(1-p_{ij})\,\E[K(x|\t)]+p_{ij}g_{ji}(x)\,.   
\end{align}
Substituting back to equation (\ref{fifj1}) one obtains
$$
\E[f_i(x)f_j(x)]=(1-p_{ij}p_{ji})\,\E[K(x|\t)]^2+p_{ij}p_{ji}\,\E\left[g_{ji}(x)^2\right].
$$
Using lemma $1$, the last equation becomes
$$
\E[f_i(x)f_j(x)]={\l_{ji}p_{ji}p_{ij}\over 2-\l_{ji}}\left\{\E[K(x|\t)^2]-\E[K(x|\t)]^2\right\}+\E[K(x|\t)]^2,
$$
or that
$$
{\rm Cov}(f_j(x),f_i(x))\, =\, {\l_{ji}p_{ji}\,p_{ij}\over 2-\l_{ji}}{\rm Var}(K(x|\t)).
$$
The desired result, comes from the fact that
\begin{align}
\nonumber
{\rm Var}\left(\int_\Theta K(x|\t)\G_{ji}(d\t)\right) 
       & =
\left\{{\l_{ji}\over 2-\l_{ji}}\E[K(x|\t)^2]+{2(1-\l_{ji})\over 2-\l_{ji}}\E[K(x|\t)]^2\right\}-\E[K(x|\t)]^2\\
\nonumber
       & = {\l_{ji}\over 2-\l_{ji}}\left(\E[K(x|\t)^2]-\E[K(x|\t)]^2\right).
\end{align}
\hfill $\square$

\medskip\noindent{\bf Proof of Proposition 4.}\\
(1.) From equation (\ref{VarPDSBP}) and proposition 3, we have that
$$
{\rm Var}(f_j^{\cal G}(x))
=  {\rm Var}\left(\sum_{l=1}^m p_{jl}g_{jl}^{\cal G}(x)\right)
=\sum_{l=1}^m {p_{ji}^2\lambda_{ji}\over 2-\lambda_{ji}}{\rm Var}(K(x|\t)).
$$
Normalizing the covariance in equation (\ref{CovPDSBP}) with the associated
standard deviations, yields
\begin{equation}
\label{CorrG}
{\rm Corr}(f_j^{\cal G}(x),f_i^{\cal G}(x))  = {\lambda_{ji}p_{ji}p_{ij}\over 2-\lambda_{ji}}
\left(\sum_{l=1}^m\sum_{r=1}^m  {p_{jl}^2 p_{ir}^2\lambda_{jl}\lambda_{ir}\over (2-\lambda_{jl})(2-\lambda_{ir})}\right)^{-1/2}.
\end{equation}
Similarly, from proposition 1 in Hatjispyros et al. (2011), it is that
$$
{\rm Var}(f_j^{\cal D}(x))
=\sum_{l=1}^m {p_{ji}^2\over 1+c_{ji}}{\rm Var}(K(x|\t)),
$$
and
\begin{equation}
\label{CorrD}
{\rm Corr}(f_j^{\cal D}(x),f_i^{\cal D}(x))  = {p_{ji}p_{ij}\over 1+c_{ji}}
\left(\sum_{l=1}^m\sum_{r=1}^m  {p_{jl}^2 p_{ir}^2\lambda_{jl}\lambda_{ir}\over (1+c_{jl})(1+c_{ir})}\right)^{-1/2}.
\end{equation}

\medskip\noindent (2.)
When $\lambda_{ji}=\lambda$ and $c_{ji}=c$ for all $1\le j\le i\le m$, from equations (\ref{CorrG})
and (\ref{CorrD}), it is clear that 
$$
{\rm Corr}(f_j^{\cal G}(x),f_i^{\cal G}(x))={\rm Corr}(f_j^{\cal D}(x),f_i^{\cal D}(x))
 = p_{ji}p_{ij}\left( \sum_{l=1}^m \sum_{r=1}^m  p_{jl}^2 p_{ir}^2 \right)^{-1/2}.
$$


\vspace{0.2in} \noindent{\bf Appendix B}

\vspace{0.2in} \noindent{\bf 1. Sampling of the concentrations masses for the rPDDP model.}

\noindent
In this case, the random densities $(f_j)$ are represented as finite mixtures of the 
DP mixtures $g_{jl}(x|\P_{jl})$, where $\P_{jl}\sim{\cal DP}(c_{jl},P_0)$. 
We randomize the concentrations by letting $c_{jl}\sim{\cal G}(a_{jl},b_{jl})$.
Following West (1992) we have the following two specific cases:

\medskip\noindent {\bf A.}  For $j=l$, the posterior $c_{jj}$'s will be 
affected only by the size of the data set ${\boldsymbol x}_j$ and the number of unique clusters for which 
$\d_{ji}={\bf e}_j$. Letting 
$$
\rho_{jj}=\#\{d_{jj}:\d_{ji}={\bf e}_j,1\le i\le n_j\},
$$ 
we have
\begin{align}
\nonumber
 & \b\sim{\cal B}e(c_{jj}+1, n_{j})\\
\nonumber
 & c_{jj}\,|\,\b,\rho_{jj}\, \sim\, \pi_\b\,{\cal G}(a_{jj}+\rho_{jj}, b_{jj}-\log\b) +
         (1-\pi_\b)\,{\cal G}(a_{jj}+\rho_{jj}-1, b_{jj}-\log\b)
\end{align}
with the weights $\pi_\b$ satisfying
$\frac{\pi_\b}{1-\pi_\b}=\frac{a_{jj}+\rho_{jj}-1}{n_{j}(b_{jj}-\log\b)}$.

\medskip\noindent {\bf B.} For $j\neq l$, the posterior $c_{jl}$'s will be affected by the size 
of the data sets ${\boldsymbol x}_j$ and ${\boldsymbol x}_l$ and the cumulative number 
of unique clusters $d_{ji}$ for which $\d_{ji}={\bf e}_l$ and the
unique clusters $d_{li}$ for which $\d_{li}={\bf e}_j$. Letting
$$
\rho_{jl}=\#\{d_{ji}:\d_{ji}={\bf e}_l,1\le i\le n_j\}+
          \#\{d_{li}:\d_{li}={\bf e}_j,1\le i\le n_l\},
$$ 
it is that
\begin{align}
\nonumber
 & \b\sim{\cal B}e(c_{jl}+1, n_{j}+n_{l})\\ 
\nonumber
 & c_{jl}\,|\,\b,\rho_{jl}\,\sim\,\pi_\b\,\mathcal{G}(a_{jl}+\rho_{jl}, b_{jl}-\log\b) +
         (1-\pi_\b)\,\mathcal{G}(a_{jl}+\rho_{jl}-1, b_{jl}-\log\b),
\end{align}
with the weights $\pi_\b$ satisfying 
$\frac{\pi_\b}{1-\pi_\b}=\frac{a_{jl}+\rho_{jl}-1}{(n_{j}+n_{l})(b_{jl}-\log\b)}$.

Bear in mind that $\rho_{jl}=0$ is always a possibility, so that we impose $a_{jl}>1$.

\vspace{0.2in} \noindent{\bf 2. Sampling of the geometric probabilities for the PDGSBP model.}

\noindent
In this section we provide the full conditionals for the geometric probabilities $\lambda_{jl}$ 
under beta conjugate and transformed gamma nonconjugate priors.
We let 
$$
S_{jl} =\sum_{i=1}^{n_j}{\cal I}(\d_{ji}={\bf e}_l)\quad{\rm and}\quad
S_{jl}'=\sum_{i=1}^{n_j}{\cal I}(\d_{ji}={\bf e}_l)(N_{ji}-1).
$$
 
\medskip\noindent 
{\bf A.} For the choice of prior $\lambda_{jl}\sim{\cal B}e(a_{jl},b_{jl})$, 
for $l=j$ it is that
$$
f(\l_{jj}|\cdots)={\cal B}e(\l_{jl}| a_{jj} + 2 S_{jj}, b_{jj} + S_{jj}'),
$$
also, for $l\neq j$ we have
$$
f(\l_{jl}|\cdots) = {\cal B}e(\l_{jl}| a_{jl} + 2(S_{jl}+S_{lj}), b_{jl} + S_{jl}' + S_{lj}').
$$

\medskip\noindent 
{\bf B.} For the choice of prior  $\l_{jl}\sim{\cal TG}(a_{jl},b_{jl})$, for $l=j$ it is that
$$
f(\l_{jj}|\ldots) \propto \l_{jj}^{2S_{jj} - a_{jj}-1}(1-\l_{jj})^{S_{jj}'+ a_{jj}-1}e^{-b_{jj}/\l_{jj}}\,{\cal I}(0<\l_{jj}<1).
$$
To sample from this density, we include the positive auxiliary random variables $\nu_1$ and $\nu_2$ such that
$$
f(\l_{jj},\nu_1,\nu_2|\cdots) \propto 
\l_{jj}^{2S_{jj} - a_{jj}-1}{\cal I}\left(\nu_1<(1-\l_{jj})^{S_{jj}'+ a_{jj}-1}\right) 
{\cal I}\left(\nu_2<e^{-b_{jj}/\l_{jj}}\right){\cal I}(0<\l_{jj}<1).
$$
The full conditionals for $\nu_1,\nu_2$ are uniforms
$$
f(\nu_1|\cdots) ={\cal U}\left(\nu_1|0, (1-\l_{jj})^{S_{jj}'+ a_{jj}-1}\right)\quad{\rm and}\quad
f(\nu_2|\cdots) ={\cal U}\left(\nu_2|0, e^{-b_{jj}/\l_{jj}}\right),
$$
and the new full conditional for $\lambda_{jj}$ becomes
$$
f(\l_{jj}|\nu_1,\nu_2,\ldots) \propto \l_{jj}^{2S_{jj} - a_{jj}-1}
 \begin{cases} 
      \hfill {\cal I}\left(-{b_{jj}\over\log\nu_2}<\l_{jj}<1-\nu_1^{1/L_{jj}}\right)    \hfill & L_{jj}\ge 0 \\
      \hfill {\cal I}\left(\max\left\{-{b_{jj}\over\log \nu_2},1-\nu_1^{1/L_{jj}}\right\}<\l_{jj}<1\right) \hfill & L_{jj}<0, \\
  \end{cases}
$$
where we have set $L_{jj}=S_{jj}'+ a_{jj}-1$.
We can sample from this density using the inverse cumulative distribution function technique.
Also, for $l\neq j$ we apply the same embedded Gibbs sampling technique to the full conditional 
density
$$
f(\l_{jl}|\cdots)\propto\l_{jl}^{2(S_{jl}+S_{lj})-a_{jl}-1}(1-\l_{jl})^{S_{jl}'+S_{lj}'+ 
a_{jl}-1}e^{-b_{jl}/\l_{jl}}\,{\cal I}(0<\l_{jl}<1).
$$


\vspace{0.2in} \noindent{\bf References.}

\begin{description}

\item {\sc Bulla, P., Muliere, P. and Walker, S.G. (2009)}.
A Bayesian nonparametric estimator of a multivariate survival function.
{\sl Journal of Statistical Planning and Inference} \textbf{139}, 3639--3648. 

\item {\sc De Iorio, M., M\"uller, P., Rosner, G.L. and MacEachern, S.N. (2004)}.
An ANOVA model for dependent random measures.
{\sl Journal of the American Statistical Association} \textbf{99}, 205--215. 

\item {\sc Dunson, D.B. and Park, J.H. (2008)}.
Kernel stick--breaking processes.
{\sl Biometrika} \textbf{95}, 307--323.

\item {\sc Ferguson, T.S. (1973)}.
A Bayesian analysis of some nonparametric problems.
{\sl Annals of Statistics} \textbf{1}, 209--230.  

\item {\sc Fuentes--Garcia, R., Mena, R.H., Walker, S.G. (2009)}.
A nonparametric dependent process for Bayesian regression
{\sl Statistics and Probability Letters} \textbf{79}, 1112--1119.  

\item {\sc Fuentes--Garcia, R., Mena, R.H., Walker, S.G. (2010)}.
A new Bayesian nonparametric mixture model.
{\sl Comm.Statist.Simul.Comput} \textbf{39}, 669--682.  

\item {\sc Griffin, J.E. and Steel, M.F.J. (2006)}.
Order--based dependent Dirichlet processes.
{\sl Journal of the American Statistical Association} \textbf{101}, 179--194.

\item {\sc Griffin, J.E., Kolossiatis, M. and Steel, M.F.J. (2013)}.
Comparing distributions by using dependent normalized ranom--measure mixtures.
{\sl Journal of the Royal Statistical Society, Series B} \textbf{75}, 499--529.

\item {\sc Hatjispyros, S.J., Nicoleris, T. and Walker, S.G. (2011)}.
Dependent mixtures of Dirichlet processes.
{\sl Computational Statistics and Data Analysis} \textbf{55}, 2011--2025.

\item {\sc Hatjispyros, S.J., Nicoleris, T. and Walker, S.G. (2016a)}.
Dependent random density functions with common atoms and pairwise dependence.
{\sl Computational Statistics and Data Analysis} \textbf{101}, 236--249.

\item {\sc Hatjispyros, S.J., Nicoleris, T. and Walker, S.G. (2016b)}.
Bayesian nonparametric density estimation under length bias.
{\sl Communications in Statistics}\\ DOI: 10.1080/03610918.2016.1263735

\item {\sc Lijoi, A., Nipoti, B. and Pr\"uenster, I. (2014a)}. 
Bayesian inference with dependent normalized completely random measures. 
{\sl Bernoulli}, {\bf 20}, 1260--1291.

\item {\sc Lijoi, A., Nipoti, B. and Pr\"uenster, I. (2014b)}.
Dependent mixture models: clustering and borrowing information.
{\sl Computational Statistics and Data Analysis} \textbf{71}, 17--433.

\item {\sc Kolossiatis, M., Griffin, J.E. and Steel, M.F.J. (2013)}.
On Bayesian nonparametric modelling of two correlated distributions.
{\sl Statistics and Computing} \textbf{23}, 1--15.

\item {\sc Lo, A.Y. (1984)}.
On a class of Bayesian nonparametric estimates I. Density estimates.
{\sl Annals of Statistics} \textbf{12}, 351--357.

\item {\sc MacEachern, S.N. (1999)}.
Dependent nonparametric processes. In
{\sl ``Proceedings of the Section on Bayesian Statistical Science''} pp. 50-55. American Statistical Association.

\item {\sc M\"uller, P., Quintana, F., and Rosner, G., (2004)}.
A method for combining inference across related nonparametric Bayesian models.
{\sl Journal of the Royal Statistical Society, Series B} \textbf{66}, 735--749.

\item {\sc Mena, R.H., Ruggiero, M. and Walker, S.G. (2011)}.
Geometric stick--breaking processes for continuous--time Bayesian nonparametric modeling.
{\sl Journal of Statistical Planning and Inference} \textbf{141} (9), 3217--3230.

\item {\sc Sethuraman, J. (1994)}.
A constructive definition of Dirichlet priors.
{\sl Statistica Sinica} \textbf{4} 639--650.

\item {\sc Walker, S.G. (2007)}.
Sampling the Dirichlet mixture model with slices
{\sl Communications in Statistics} \textbf{36} 45--54.

\item {\sc West, M. (1992)}.
Hyperparameter estimation in Dirichlet process mixture models.
{\sl Technical report} \textbf{92-A03}, Duke University, ISDS.

\end{description}

\end{document}